\documentclass[twocolumn,showpacs,preprintnumbers,amsmath,amssymb]{revtex4}

\usepackage{graphicx}
\usepackage{dcolumn}
\usepackage{bm}

\begin{document}

\preprint{APS/123-QED}

\title{Polarizability measurements in a molecule near-field interferometer}
\author{Martin Berninger}
\author{Andr\'{e} Stefanov}
\author{Sarayut Deachapunya}
\author{Markus Arndt}
\email{markus.arndt@univie.ac.at}
\affiliation{Fakult\"{a}t f\"{u}r Physik der Universit\"{a}t Wien, Boltzmanngasse 5, A--1090 Wien\\}%

\date{\today}

\begin{abstract}
We apply near-field matter-wave interferometry to determine the
absolute scalar polarizability of the fullerenes C$_{60}$ and
C$_{70}$. A key feature of our experiment is the combination of
good transmission and high spatial resolution, gained by wide
molecular beams passing through sub-micron gratings. This allows
to significantly facilitate the observation of field-dependent
beam shifts. We thus measure the polarizability to be $\alpha=88.9
\pm 0.9 \pm 5.1 \, \rm \AA^{3}$ for C$_{60}$ and to $\alpha =
108.5 \pm 2.0 \pm 6.2 \, \rm \AA^{3}$ for C$_{70}$.
\end{abstract}

\maketitle
Knowing the scalar polarizability  of
a molecule is of importance in various areas of physics.
 The polarizability may provide one of several parameters in the
description of molecular shapes~\cite{Bonin1997a}. Even for the
structurally simple molecules $\textrm{C}_{60}$ and
$\textrm{C}_{70}$, various theoretical models have been competing
with each other~\cite{Compagnon2001c}.

The molecular response to external electric fields is also the
handle for slowing~\cite{Maddi1999a,Bethlem1999a,Fulton2004a}
cooling or trapping \cite{Takekoshi1995a} neutral particles in
switched electric fields or off-resonant laser beams.

It also turns out to
be crucial for matter-wave interferometry. The
van der Waals or Casimir-Polder  potentials acting between the
traversing molecules and the diffraction gratings have an enormous
influence on the details of the interference fringes~\cite{Grisenti1999a,Brezger2002a}.
And this will generally grow with the the size of the particles.

Also the small-angle scattering cross section
between different molecules is related to their polarizability. And this
enters, for instance, experimental considerations through the definition of the vacuum conditions
which are required for avoiding collisional decoherence in
interferometry~\cite{Hornberger2003a}.

Finally, precise $\alpha$-measurements are also stimulated by the
insight that future molecule interferometers will also exploit optical
gratings~\cite{Gould1986a,Nairz2001a,Brezger2003a}. There
the polarizability is needed to determine the phase shift caused
by the interaction between the induced dipole moment and the
electric field of the diffracting laser beam.

Various experimental approaches are conceivable to determine
$\alpha$ with high precision~\cite{Compagnon2001c}. In solids or
liquids, the polarizability can be obtained through a measurement
of the dielectric constant and the index of
refraction~\cite{Bonin1997a}. But in particular for the comparison
with theoretical models it is desirable to get information about
the isolated particles \cite{Simons2005a}.

Polarizabilities of dilute molecular beams are usually
characterized in Stark-deflection experiments, in which the
neutral molecules are traversing a homogeneous electrostatic force
field. The polarizability can then be determined by measuring the
lateral shift of the molecular beam due to this field. In most
experiments the resulting deflections are much smaller than the
width of the molecular beam profile~\cite{Compagnon2001c} but
still such experiments can typically reach an accuracy of about
10\% ~\cite{Antoine1999a}. One may also consider the use of
electric field gradients to induce a longitudinal velocity shift
and corresponding time-of-flight delay in a molecular fountain
arrangement~\cite{Amini2003a}.

For atoms, very precise polarizability values were also obtained
using far-field matter-wave interferometry: the atomic wave
function can be coherently split and guided through two spatially
separated electric fields. The different phases imprinted on the
atom through the interaction between $\alpha$ and electric field
$E$ along the two different paths shows up as a fringe shift that
can be well resolved. This allowed a very accurate determination
of ground state polarizabilities for Sodium~\cite{Ekstrom1995a}
and Lithium~\cite{Miffre2006a}.

Efficient methods are also highly desirable for metrology on large molecules,
in particular since their $\alpha/m$-values may be significantly smaller than those of alkali
atoms. The use of far-field interferometry
would, however, require beam sources of very high spatial coherence. In
the absence of efficient cooling and phase-space compression
methods, this can only be achieved by strictly collimating the
molecular beam. We have demonstrated far-field diffraction successfully in our
earlier work for large carbon molecules~\cite{Arndt1999a} but the
limited flux of almost all other large particles would prohibit such a
strict spatial selection.
While classical Stark deflection usually
provides a high flux with a low spatial resolution,
far-field interferometers trade their high-resolution in for a strongly restricted molecular
transmission.

Here we report on a solution of this problem, which is based on a
near-field Talbot-Lau interferometer (TLI).
It combines high spatial resolution with high molecular throughput and
in can be generalized to objects of arbitrary mass --- in the limit
of classical Moir\'{e} deflectometry.

\begin{figure}[h]
\begin{center}
\includegraphics[width=8.5cm, keepaspectratio=true]{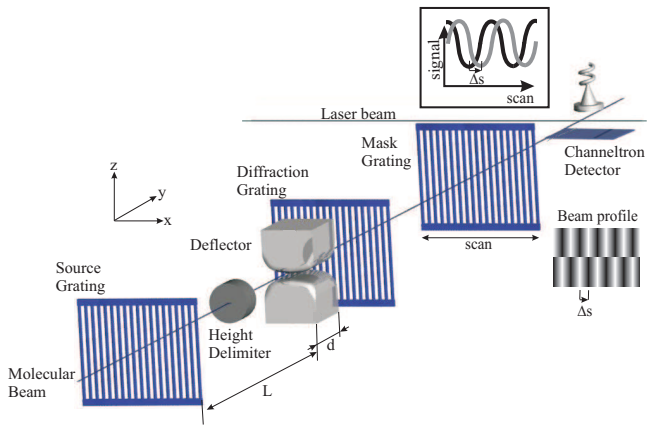}
\caption{Experimental set-up: The three gratings are used for
coherence preparation, diffraction and detection. Path-dependent
matter-wave phase shifts in the external field lead to a beam
deflection at the position of the mask grating. The lateral shift
$\Delta s$ of the interference fringes at the detector is directly
proportional to the scalar molecular polarizability $\alpha$.}
\label{fig:Deflektometer}
\end{center}
\end{figure}

A Talbot-Lau interferometer uses near-field wave effects, generally for the lensless imaging of
periodic nanostructures~\cite{Patorski1989a}.
It has already been described earlier in the context of atom~\cite{Clauser1992a,Clauser1997b}
and molecule~\cite{Brezger2002a,Hackermuller2003a} interferometry.
In our present apparatus we employ a TLI composed of three gold gratings with
a grating period of g=991\,nm and slit openings of about 450\,nm width.

The first grating prepares the required spatial coherence to
obtain matter wave diffraction at the second grating, which then
results in a regular molecular density pattern, immediately in
front of the mask grating. The interference fringes, whose
periodicity equals the gratings constant, can be revealed by
scanning the third grating laterally. The resulting modulation of
the molecular flux behind this setup allows to visualize the
molecular nano-fringe system. The molecular beam width in a
TL-interferometer may be wider than 1\,mm, allowing a high
throughput of molecules while the resolution of the fringe system
can be better than 15\,nm.

We now insert an electrostatic deflector into the TLI (see
Fig.~\ref{fig:Deflektometer}), which creates a homogeneous force
field $\mathbf{F} =\alpha (\mathbf{E} \nabla E)$ changing the
momentum of the molecules in proportion to the applied electric
field $\mathbf{E} $, its gradient  $\nabla \mathbf{E}$ and to the
polarizability $\alpha$. In the proper quantum picture, the
electric field adds a path-dependent phase shift on the molecular
matter wave. In both descriptions, classical and quantum, we
obtain a fringe shift:
\begin{equation}\label{equ:shift}
\Delta s_{x} (\alpha, v)= \alpha \frac{(\mathbf{E} \nabla)
E_{x}}{m} \frac{d}{v^{2}} \bigg( \frac{d}{2} + L \bigg)
\end{equation}
where $m$ corresponds to the molecular mass, $v$ the velocity
along the direction of the beam and $L$ the distance between the
source grating and the deflector. The front edge of the deflector
(length $d$) is positioned at $L = (26.6 \pm 0.1 )\, \textrm{cm}$
behind the first grating. The deflector provides a constant force
with deviations of $0.5 \%$ along the x-axis over the entire
molecular beam diameter.

\begin{figure}[h]
\begin{center}
\includegraphics[width=0.9\columnwidth]{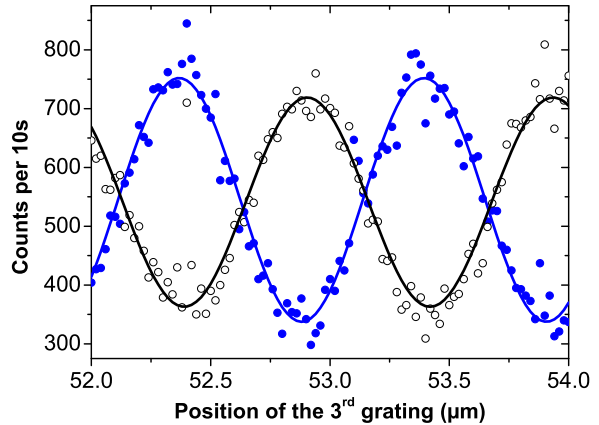}
\caption{Deflection of a C$_{60}$ beam with $\bar{v}=117$\,m/s and
$\sigma_{v} = 8\,\%$. A phase shift of $\Delta \phi =\pi$ is
obtained at a voltage of 6\,kV (full circles). The open circles
represent the reference at U=0\,kV.}\label{fig:shift}
\end{center}
\end{figure}

Two interference recordings at two different deflection voltages
would be sufficient to determine the molecular polarizability, if
all parameters were known with arbitrary precision.
Figure~\ref{fig:shift} shows a fringe system of C$_{60}$ with a
deflection voltage of 0\,kV and 6\,kV. The field-dependent shifts
can be nicely resolved and measured with high accuracy. In order
to obtain additional statistical information and to assess the
accuracy of the various contributing parameters, we repeat the
measurements for different velocity distributions and for
different deflection voltages.
We employ a gravitational velocity
selection~\cite{Arndt2001a,Brezger2002a} and chose the velocity by
moving the source to its appropriate vertical position. An
interference pattern is then recorded by displacing the third
grating in steps of 20\,nm over about three fringe periods and by
repeating this scan for all voltages within 3--15\,kV in steps of
1\,kV.

\begin{figure}[h]
\begin{center}
\includegraphics[width=1\columnwidth]{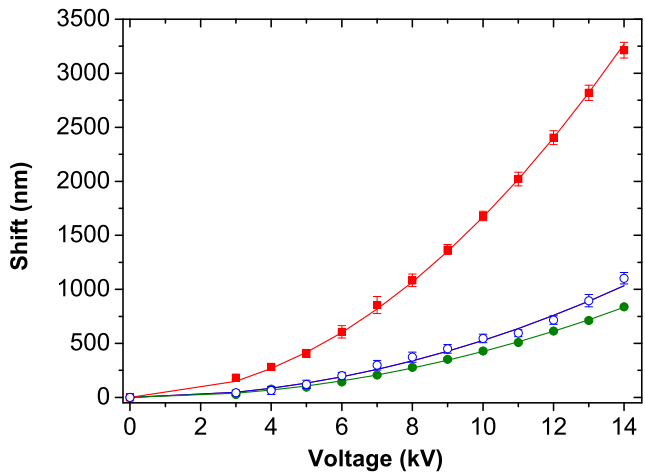}
\caption{The  Stark-deflection of the C$_{70}$ fringes follows
precisely the quadratic voltage dependence $\Delta s \propto
U^2/v^2$ of Eq.\,\ref{equ:shift}: full circles: $\bar{v}$ = 199 m/s,
$\sigma_{v}= 16 \%$, open circles $\bar{v}$ = 173 m/s, $\sigma_{v}=
13 \%$, full squares: $\bar{v}$ = 109 m/s, $\sigma_{v}= 7
\%$.}\label{fig:Shift_versus_Voltage}
\end{center}
\end{figure}

To monitor and numerically compensate for drifts, an additional
reference point (with U=0\,kV) is always included before and
after each high-voltage deflection scan. From the interference
curves thus obtained we extract the voltage dependence of the
experimental fringe shift (Fig.~\ref{fig:Shift_versus_Voltage}).
The observed fringe shift is influenced by the details of the
molecular velocity distribution in three ways. Firstly, slow
molecules will acquire a larger deflection in the external field
gradient than fast ones (Eq.~\ref{equ:shift}). Secondly,
different velocity classes are associated with different
visibilities. This is an important and desired feature of the
Talbot-Lau arrangement, as it allows to prove the quantum
wave-nature of material objects~\cite{Brezger2002a}. Thirdly, the
van der Waals interaction with the grating walls adds a
dispersive phase shift~\cite{Grisenti1999a,Brezger2002a}.

The expected signal as a function of the scanning grating position
$x$ is proportional to:

\begin{eqnarray}
1+\bar{V}_{th}(\alpha )\cos \big[\frac{2\pi}{g}\big(x-\Delta
s_{th}(\alpha)\big)\big] \qquad \qquad&&\nonumber\\
\equiv \int \textrm{d}v f_{\bar{v},\sigma_{v}}(v)\bigg(1+ V(v)
\cos \big[ \frac{2\pi}{g} \big( x- \Delta s_{x} (\alpha, v) \big)
\big]\bigg)
\end{eqnarray}

where the velocity distribution function
$f_{\bar{v},\sigma_{v}}(v)$ is experimentally determined for
several source settings, i.e. different mean velocities $\bar{v}$.
The fringe visibility function $V(v)$ cannot be measured directly
because of the finite width of the velocity distribution
$\sigma_{v}$. It is extracted from the measured visibilities
$V(\bar{v},\sigma_v)$ using

\begin{equation}
V(\bar{v},\sigma_v)=\int \textrm{d}v f_{\bar{v},\sigma_{v}}(v)
V(v).
\end{equation}

The polarizability $\alpha$ is then obtained by fitting the
experimentally observed shift $\Delta s_{\textrm{exp}}$ with the
theoretical shift $\Delta s_{\textrm{th}}(\alpha)$.

\begin{figure}[h]
\begin{center}
\includegraphics[width=1\columnwidth]{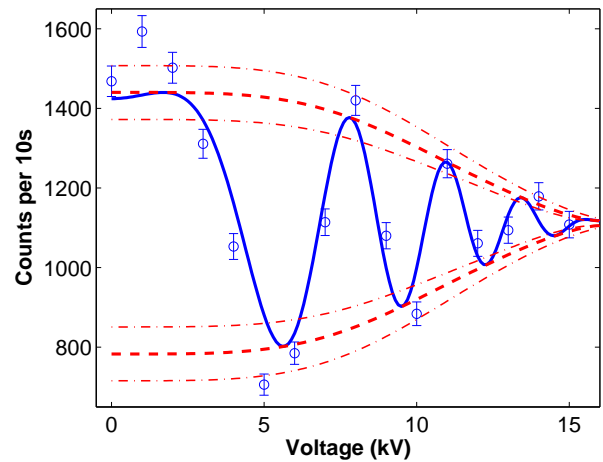}
\caption{Signal as a function of the deflection voltage for a
fixed position of the mask grating. The open circles are data
points of C$_{70}$ at a velocity of $\bar{v} = 103.2 m/s,
\sigma_{v}= 7 \%$. The solid line is the theoretical expectation
for $\alpha= 108.5 \, \AA^{3}$. The dashed line is the envelope
curve of the visibility function. The dashed dotted lines
illustrate upper and lower bounds resulting from small variations
of the velocity distribution: $\Delta \bar{v} = \pm 1\%, \Delta
\sigma_{v}= \pm 1\%$ and typical uncertainties of the visibility:
$\Delta V(v) = \pm 3\%$. }\label{fig:Visibility_Versus_Voltage}
\end{center}
\end{figure}

We limit the maximum deflection voltage in the experiment to
15\,kV since the fringes of slow (109\,m/s) fullerenes are then
already shifted by three entire periods
(Fig.~\ref{fig:Shift_versus_Voltage}) and dephasing effects
become important. Molecules of different velocities experience a
different Stark deflection, and the summation over the
corresponding molecule interference patterns in the detector will
reduce the total fringe visibility  for sufficiently large
velocity spreads $\sigma_v$ and with increasing voltage (see
Fig.~\ref{fig:Visibility_Versus_Voltage}). In our experiments
$\sigma_v (1/e^2)$ ranges between 7\,\% and 16\,\% for molecules
between 100\,m/s and 200\,m/s, respectively.

When all grating positions are fixed, the scanning deflection
voltage can be used to sweep the molecular density pattern across
the third grating (Fig.~\ref{fig:Visibility_Versus_Voltage}). The
data points are well described using the implicit functions of the
voltage for the expected visibility $\bar{V}_{th}(\alpha)$ and
shift $\Delta s_{th}(\alpha)$. It is interesting to note, that
not only the shift but also the voltage dependent dephasing, which
leads to a rather sharp decrease in the interference contrast,
may actually be used as tools for metrology. The fringe shifts
are, however, less sensitive to additional random perturbations
-- such as vibrations, thermal radiation or collisions -- than the
visibility, and the values given below are therefore based on the
displacement measurements.

We find a polarizability of $88.9 \pm 0.9 \pm 5.1 \, \rm \AA^{3}$
for C$_{60}$  and  $108.5 \pm 2.0 \pm 6.2 \, \rm \AA^{3}$ for
C$_{70}$, including the statistical (first) and systematic
(second) error. The relative polarizability is then $\alpha
(\textrm{C}_{70}) / \alpha (\textrm{C}_{60}) = 1.22 \pm 0.03$.
Most of the systematic uncertainties, such as for instance field
inhomogeneities, cancel out in the determination of the relative
values.

The statistical uncertainty of 2\,\%, given above, refers to the
standard deviation in the determination of $\alpha$ for different
velocities. If all data were purely shot noise limited, we would
expect an uncertainty as small as 0.1\,\%.

The uncertainty in the measurement of the velocity amounts to
$1\,\%$, the voltages are known to $0.5\,\%$. The lateral
resolution of our interferometer is currently good to within
$\sim15\,\textrm{nm}$. This amounts to less than two percent of
the maximum fringe shift and it corresponds to a lateral force
sensitivity of better than $10^{-14}$\,pN.

The properties of the electric field are characterized using a
finite element code~\cite{Femlab2004a} which finds a value of
$(\mathbf{E} \nabla) E_{x} = (1.45 \pm 0.08 ) \cdot 10^{14} \,
\textrm{V}^{2}/\textrm{m}^{3}$ for a voltage of 10\,kV. The same
code is used  to determine the effective electrode length
$d_{\textrm{eff}} = (4.73 \pm 0.1) \, \textrm{cm}$ which replaces
the real electrode length because of edge effects. The remaining
uncertainties comprise the accuracy in the measurement of the
electrode's contour and separation. All independent systematic
uncertainties add up to a total of $5.7\,\%$.

In conclusion, we have demonstrated that a near-field matter-wave
interferometer is a promising tool for sensitively measuring the
scalar polarizability. The three-grating design allows us to
combine high spatial resolution and high molecular flux.

As mentioned before, the velocity spread is a
critical parameter, since it enters quadratically in
Eq.~\ref{equ:shift}. However, in particular in pulsed
molecular beam experiments, based on laser desorption and on pulsed photo-ionization
the velocity selection can reach values as small as $\Delta v/v\sim 0.1\,\%$
at acceptable signal intensities~\cite{Marksteiner2006a},
even if the entire beam has still a finite temperature.
This way it is already possible to generate neutral beams of fullerenes,
complex biomolecules~\cite{Meijer1990a,Antoine1999a} up to polypeptides
composed of 15 amino acids~\cite{Marksteiner2006a}.

The extension of our present setup to pulsed sources and detectors
is particularly interesting for instance for investigating the
temperature dependence of the molecular polarizability. But also
magnetic moments or triplet lifetimes of fullerenes should thus be
accessible.

In biomolecules, the conformational variation of polarizabilities
and electric dipole moments will be of interest. If the molecules
exhibit a finite electric or magnetic dipole moment, the random
orientation of cold molecules will lead to a decrease of the
visibility instead of a simple fringe shift \cite{Antoine2002a}.
As mentioned above, the amount of dephasing will then allow the
extraction of a numerical value for the moments. For hot
molecular beams which possess a permanent electric dipole moment
the system behaves as an object with only an induced
polarizability \cite{Moro2006a}. Therefore, dipole moments could
already be measured in our current setup by only replacing the
source.

With respect to far-field interference methods the near-field scheme
can be much extrapolated towards significantly more complex particles.
And even in its classical limit, as a multiplexing Stark-deflectometer
it will still combine high resolution with high transmission.
Finally, it is interesting to note, that our apparatus may also be regarded as a
fast switch for molecular beams. The fine-structure imprinted on the beam by its
transmission through the first two gratings allows us to lower and raise the
molecular transmission with kHz-frequencies. In dedicated experiments it should be
possible to reach a modulation amplitude of close to 100\%.

\begin{acknowledgments}
Our work is supported by the Austrian FWF through SFBF1505,
STARTY177-2 and the European Commission under contract
HPRN-CT-2002-00309. S. Deachapunya is supported through a Royal
Thai Government scholarship. We thank H. Ulbricht for fruitful
discussions.
\end{acknowledgments}

\end{document}